\documentclass[11pt]{article}
\usepackage{moriond_ke,epsfig}

\def\Journal#1#2#3#4{{#1} {\bf #2}, #3 (#4)}

\def\MPLA{{\em Mod. Phys. Lett.}  A}

\def\NIMA{{\em Nucl. Instrum. Methods} A}
\def\NPB{{\em Nucl. Phys.} B}
\def\PLB{{\em Phys. Lett.}  B}
\def\PRA{{\em Phys. Rev.}  A}
\def\PRL{\em Phys. Rev. Lett.}
\def\PRD{{\em Phys. Rev.} D}

\def\be{\begin{equation}}
\def\ee{\end{equation}}
\def\bea{\begin{eqnarray}}
\def\eea{\end{eqnarray}}

\begin{document}
\pagestyle{plain}
\setlength{\unitlength}{1cm}
\vspace*{4cm}
\title{The ALPS Light Shining Through a Wall Experiment - WISP Search in the Laboratory}

\author{K. Ehret - ALPS Collaboration}

\address{Deutsches Elektronen-Synchrotron DESY, Notkestra{\ss}e 85,\\
D-22607 Hamburg, Germany}

\maketitle\abstracts{
In the last years it has been realized, that extensions of the Standard Model may manifest itself also at meV energy scales. 
The low energy frontier is a rich complement to the conventional high-energy particle physics landscape.
The search for these new particles initiated experimental activities around the world.
"Light shining through a wall" (LSW) experiments search for Weakly Interacting Sub-eV Particles (WISP).  
Potential WISP candidates are axion-like particles or hidden sector photons.
The ALPS (Any Light Particle Search) experiment located at DESY in Hamburg exploits resonant laser power build-up in a large-scale optical cavity
to boost the available power for the WISP production. 
After some upgrades the experiment provides now the most stringent laboratory constraints on WISP production.
The concept, challenges and status of LSW experiments as well as their future potential are presented.}

\section{Introduction}
The Standard Model (SM) of elementary particle physics summarizes beautifully the known constituents of matter and their forces  
and describes very successfully many phenomenological observations. 
But it does not fully describe the world around us and it is generally believed that it is not the ultimate theory of matter.  
Evidence for new physics beyond the Standard Model arises from astrophysical and cosmological observations.
There are new experiments at the high-energy frontier searching for new particles beyond the SM, especially at the LHC, 
which just recently started its operation.
But there is also rising attention to the low energy frontier in particle physics, which is often explored in high precision experiments utilizing photon beams.
Especially string theory motivated extensions of the standard model predict not only new particles with masses above the 
electroweak scale ($\sim 100$ GeV) but also often a broad variety of new very light and very weakly interacting particles,
denoted as WISPs ({\bf{W}}eakly {\bf{I}}nteracting {\bf{S}}ub-eV {\bf{P}}articles), see~\cite{lef} and references therein.
WISP may interact with ordinary matter through the exchange of very massive ``mediator'' particles and thus illuminates also physics at very high energy scales.
The QCD axion originally invented to explain the CP conservation of the strong interaction
is a prime example for a WISP \cite{QCDAxion, Peccei}. 
Its mass should be below about 1~eV, implying very weak interactions with ordinary Standard Model matter \cite{Kim, PDG}.
A QCD axion with a mass below 1~meV would a perfect cold dark
matter candidate, cf.~\cite{cdm, fairbairn}.

\begin{figure}
\centerline{\includegraphics[height=60mm]{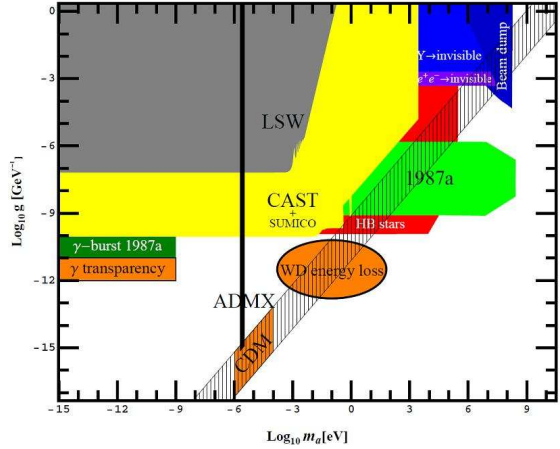}}
\caption{Constraints on axion-like-particles parameters (two
photon coupling $g$ vs. mass $m_a$ of the ALP) 
derived from cosmological and astrophysical observations and laboratory experiments \cite{Redondo:2008Patras, lef}.
\label{fig:alp_constrain}}
\end{figure}

Several astrophysical observations motivate the existence of very light axion-like particles (ALPs). 
The modeling of white dwarfs cooling benefits if an additional energy loss due to ALPs is taken into account or ALPs helps to explain the patterns of the luminosity relations of active galactic nuclei (AGN). 
Further the oscillations of photons into axion-like particles could also help to explain the high transparency of the Universe to TeV photons from AGNs or the heating of the solar corona might be attributed to an energy flow mediated by axion-like particles. Nowadays, the strongest constraints on ALPs come from astrophysical and cosmological arguments and from dedicated laboratory experiments. Fig.~\ref{fig:alp_constrain} provides an overview, cf.~\cite{lef, PDG, fairbairn, Redondo:2008Patras} and references therein.

\section{WISP Zoo and Indirect Search for New Physics at low Energies}

\begin{figure}
\includegraphics[height=30mm]{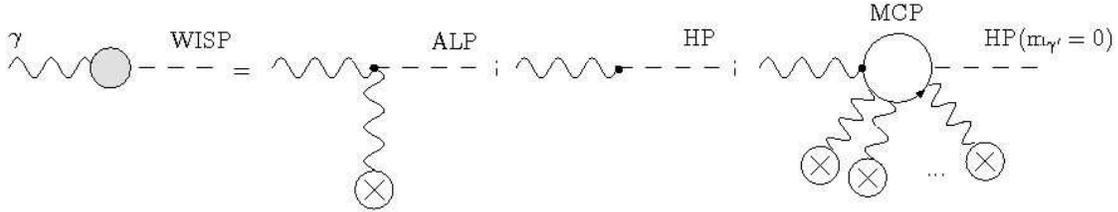}
\vspace*{-2mm}
\caption{Oscillation between photons and different types of WISP: axion-like particles (ALP), massive hidden photons (HP) and mini-charged particles (MCP) via massive hidden photons \cite{ALPSNIM, ALPSResult}.
\label{fig:WISPZoo}}
\end{figure}

One of the most exciting quests in particle physics is the search for new particles beyond the Standard Model.
Therefore it is certainly of fundamental interest, whether any of these light particles exist.
Possible candidates for WISPs are the spin-0 axion-like particles, 
which are either parity odd ALP $(0^-)$ or parity even ALP $(0^+)$, light spin 1 particles 
called ``hidden sector photons" (HP) 
or light minicharged particles (MCP).  
Fig.~\ref{fig:WISPZoo} shows the Feynman diagrams for their coupling to photons.
Photon oscillations into ALPs and MCPs require the presence of a background electromagnetic field, whereas 
oscillations into massive hidden photons occur independent of any additional external field, cf.~\cite{ALPSNIM, ALPSResult, Okun:1982xi, Raffelt:1987im, Ahlers:2007rd}.
Unfortunately, the predictions for the masses and couplings of WISPs are typically 
not precise and extensive searches in broad parameter spaces have to be performed.
Any experimental measurement which gives new indications or new limitations is highly welcome, cf.~Fig.~\ref{fig:alp_constrain}. 

QED describes photon-photon interactions via coupling to virtual electron-positron pairs.
Photons from a laser beam interacting with the virtual photons of a magnetic field 
causes the so called magnetic vacuum birefringence, a very tiny effect which changes the polarization 
of the laser beam \cite{Euler}. The corresponding QED predictions are several orders of magnitude below the sensitivity of present day experiments. The smallness of the QED effects is related to the high mass of the electron compared
to the photon energy in the eV range. Much larger effects with amplitudes well above the QCD prediction are possible, if WISPs with much smaller masses couple to photons \cite{PVLAS_Int}.
The later on not confirmed \cite{Zavattini:2007ee} observations of the PVLAS experiment in the year 2005 ~\cite{PVLAS} triggered the interest, exploration and setup of new  low energy experiments using laser beams with high photon fluxes or very good control of beam properties combined with strong electromagnetic fields. Besides PVLAS in Logarno, Italy \cite{PVLAS} indirect WISP searches are performed by OSQAR at CERN \cite{OSQAR}, BMV in Toulouse, France \cite{BMV_Pol} or Q\&A in Taiwan \cite{Q_A}. These experiments are looking for small changes of laser light polarization,
related either to vacuum magnetic birefringence or dichroism, i.e.~to dispersive or absorptive features in the propagation of polarized light along a transverse magnetic field. Dichroism related to real WISP production
rotates the polarization, whereas birefringence related to virtual WISP production causes an elliptical polarization~\cite{lef, PVLAS_Int}. Recent developments in theory, triggered by PVLAS, also creates and inspire new ideas for experimental setups. 

\section{Direct WISP Search - Light Shinning through a Wall experiments}
\begin{figure}
\centerline{\includegraphics[height=30mm]{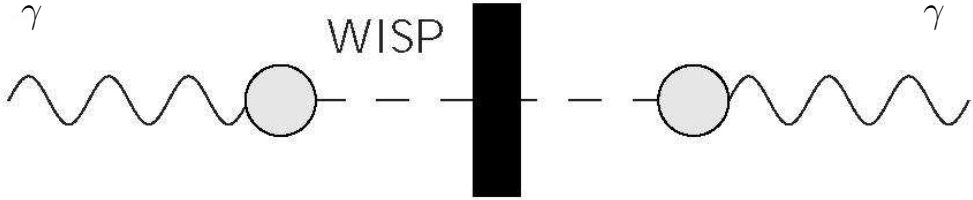}}
\begin{picture}(10,0.1)
\put(1.5,3){\Large $\gamma$}
\put(13.5,3){\Large $\gamma$}
\end{picture}
\vspace*{-8mm}
\caption{Schematic setup of a LSW experiment. Photons, typically from a laser, are shone on a light tight wall. 
Some photons may be converted 
WISPs which traverse the wall. Behind the wall some of these WISPs reconvert into ordinary photons, with the same properties as the initial photons, which are observed with a detector. 
\label{fig:WISP_lsw_scheme}}
\end{figure}

A very intriguing and simple idea to detect WISPs are ``light shining through a wall" (LSW) experiments~\cite{Okun:1982xi, Anselm:1986gz}.  
Fig.~\ref{fig:WISP_lsw_scheme} shows a sketch of such experiments. Photons, usually coming from a laser are shone on an opaque absorber, where photons are stopped, but WISPs produced in oscillations can traverse. Photons from reconverted WISPs behind the wall may be detected in a low background environment. 
This simple setup provides an enormous precision, because laser setups with a power of several hundred Watt provide more than $10^{21}$ photons per second and low noise detectors are sensitive to photon fluxes of only a few mHz. This allows to look for oscillation phenomena with transition probabilities below $10^{-24}$.

The conversion of the incident photons to axions or axion-like particles $\phi$ in the presence of a magnetic field 
is governed by the Primakoff effect~\cite{Primakoff:1951ww}. 
Behind the absorber, some of these ALPs will reconvert via the inverse-Primakoff process into photons with the initial properties. In a symmetric LSW setup the probability of the Primakoff transition $P_{\gamma \to \phi}$ is the same as for the inverse-Primakoff $P_{\phi\to \gamma}$.  
Therefore the LSW probability is just the square of $P_{\gamma \rightarrow \phi} = g^{2} B^2 E^2/(4 m_\phi^4) \cdot \sin^2 (m_\phi^2 L/(4 E))$ with $B$ the magnetic field strength, $L$ the length of the conversion region and $E$ the photon energy~\cite{lef}. Mass ($m_\phi$) and two photon coupling ($g$) of the ALPs are assumed to be uncorrelated. With $\beta_\phi$ denoting the velocity of the ALP and $q = p_\gamma - p_\phi$ it writes as:
$$ P_{\gamma \rightarrow \phi \rightarrow \gamma} = \frac{1}{16\beta^2_\phi}(g B l)^4 \left(\frac{\sin(ql/2)}{ql/2}\right)^4 $$
%
For $qL << 1$ the oscillation is coherent along the full length and the transition probability reaches its maximum $P_{\gamma \rightarrow \phi \rightarrow \gamma} = \frac{1}{16\beta^2_\phi} (g B L)^4$. The mass reach of the experiment is determined therefore by $E$ through the coherence condition $m^2_\phi < 2 \pi E / L $. 
For a larger ALP mass $m_\phi$ the momentum $p_\phi$ decreases, i.e.~the wavelength rises and runs out of phase compared to the photon wave function, causing a lower conversion probability, cf.~Fig.~\ref{fig:ALPSprospects}.
The sensitivity of LSW experiments is mainly determined by the number of incident photons, $B\cdot L$ of the magnet and the capability of the detector, cf.~Tab.~\ref{tab:alp_coupling}. The polarization of the beam allows to 
distinguish between scalar and pseudo scalar ALPs.
The BFRT experiment at Brookhaven \cite{BFRT} was the pioneering LSW experiment, it operated in the early 1990's.
The recent worldwide interest in WISP search triggered several other LSW experiments, as  BMV in Toulouse, France \cite{BMV},
GammeV at Fermilab \cite{GammeV}, LIPSS at Jefferson Lab \cite{LIPSS} or OSQAR at CERN \cite{OSQAR}. In different setups they  utilize powerful laser beams and strong magnets \cite{Lindner}. Their results are compiled in Fig.~\ref{fig:ALPSresultalps} 
and Fig.~\ref{fig:ALPSresult_hs_mcp}.

\section{The ALPS Experiment at DESY}
The {\bf ALPS} experiment located at DESY in Hamburg, initially planned for ``{\bf{A}}xion {\bf{L}}ike {\bf{P}}article {\bf{S}}earch", uses a spare superconducting HERA dipole magnet and a strong laser beam. It turned out, that the experiment has also a large sensitivity for other WISPs, so the acronym ALPS stands now more precisely for ``{\bf{A}}ny {\bf{L}}ight {\bf{P}}article {\bf{S}}earch" in a ``light shining through a wall" experiment. 
The ALPS collaboration comprises besides DESY the Albert Einstein Institute in Hannover, the Laser Zentrum Hannover and the Hamburg observatory. 

\begin{figure}
\vspace*{-2mm}
\includegraphics[height=51mm]{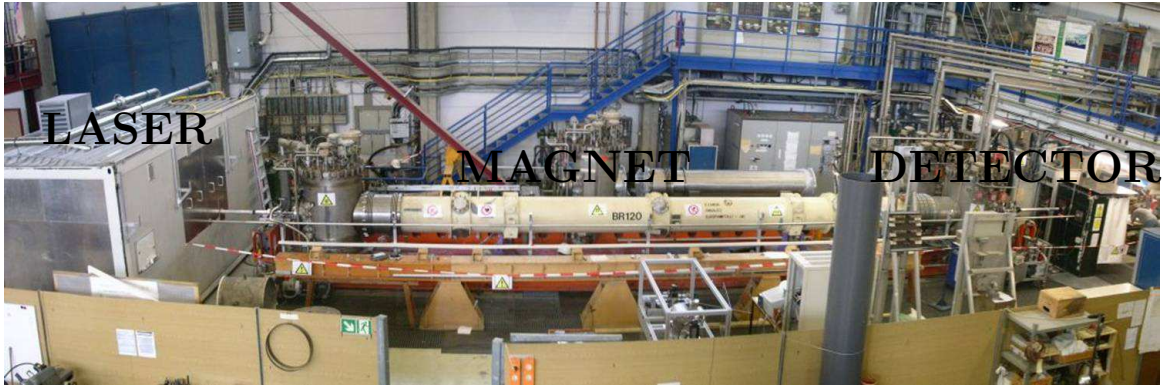}
\begin{picture}(20,0.2)
\end{picture}
\includegraphics[height=50mm]{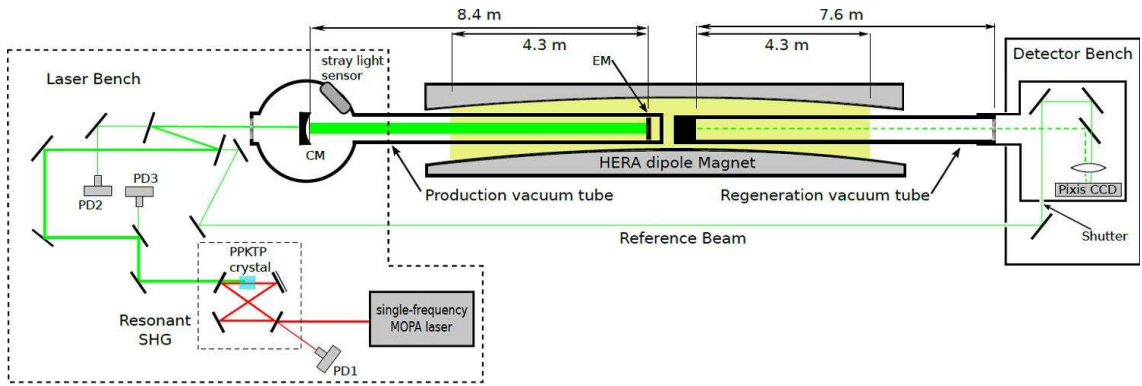}
\begin{picture}(10,0.1)
\put(0.5,9.2){\LARGE \bf LASER}
\put(6,8.7){\LARGE \bf MAGNET}
\put(11.5,8.7){\LARGE \bf DETECTOR}
\end{picture}
\vspace*{-6mm}
\caption{Photo and drawing of the ALPS experiment. PD denotes various photo detectors, CM the coupling mirror and EM the end mirror of the resonant cavity. See the text for details.
\label{fig:ALPSsetup}}
\end{figure}

Figure~\ref{fig:ALPSsetup} shows the experimental setup, which is built up around a superconducting HERA dipole with a field $B\approx 5$~T and a length of 8.42~m. The opaque wall sits in the middle of the magnet. On the left side is the laser container, providing the incident photons. On the right one sees the detector cabinet, which houses a CCD camera to detect reconverted photons. 
%
Inside the magnet the beam pipe is bent with a remaining clear aperture of only 14~mm, implying serious demands on the beam quality of the laser. The interior is insulated against the cold part of the magnet, allowing to perform the experiment at room temperature. 
Inside the dipole beam pipe two further tubes are placed, which bound the $\gamma-\phi$ conversion and reconversion regions and are operated under vacuum conditions. 
Both tubes range from either side approximately to the middle of the magnet and can be easily removed.
A removable light-tight absorber wall is mounted on the inner end of the regeneration vacuum tube while an adjustable mirror (EM) is attached to the inner side of the production vacuum tube, cf.~Fig.~\ref{fig:ALPSsetup} and Fig.~\ref{fig:ALPS_laser}

\subsection{CCD Camera as Photon Detector}
The commercial CCD camera PIXIS 1024B~\cite{PIXIS} with a pixel size of 13~$\mu$m$\times 13~\mu $m is used as detector. Operated at  $-70$  $^o$C it features a very low dark current of $0.001~e^-$/pixel/s as well as
a low read-out noise of $3.8~e^-$/pixel RMS and a very high quantum efficiency of more than 95\% for green light.
The CCD camera is mounted on a precise breadboard which contains focusing optics and a shutter for the reference beam for alignment purpose. This setup on the detector bench is attached light-tight to the detector tube, cf.~Fig.~\ref{fig:ALPSsetup}.
The beam spot is focused onto a few pixel. Groups of $3\times 3$ pixels are binned for the readout in order to lower the readout noise. Therewith nearly 90\% of the incoming light is arriving on one definned $42\times 42~\mu{\rm m}^2$ bin of 9 pixels. 
For mounting or dismounting the wall, the detector bench has to be reinstated very precisely to maintain the alignment on the pre-selected pixel. The precise and robust construction provides an accuracy for the repositioning better than $6~\mu$m. An easy control of the alignment stability is provided by the reference beam which is redirected and focused into another pixel of the CCD.
For exposures longer than $\approx \frac{1}{2}$ hour the total noise of the CCD is dominated by the dark current noise.
For much longer exposures the probability of signals generated by cosmic or radioactivity  rises, such data frames can not be used any more. Hence, usually one hour frames are taken, providing a sensitivity to a photon flux of a few mHz.


\subsection{Laser System and Resonant Photon Generation}

\begin{figure}
\includegraphics[height=64mm]{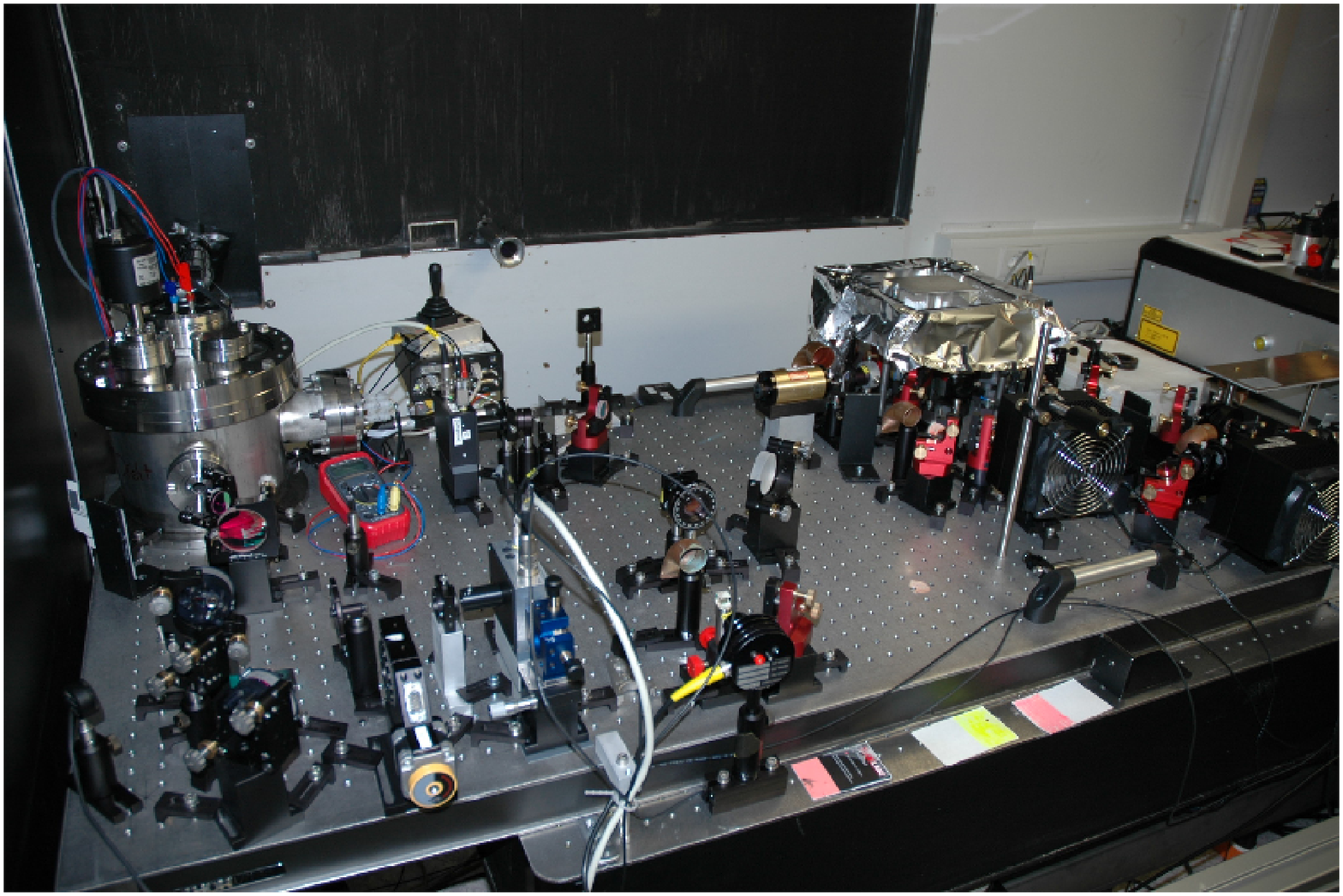}
\includegraphics[height=52mm]{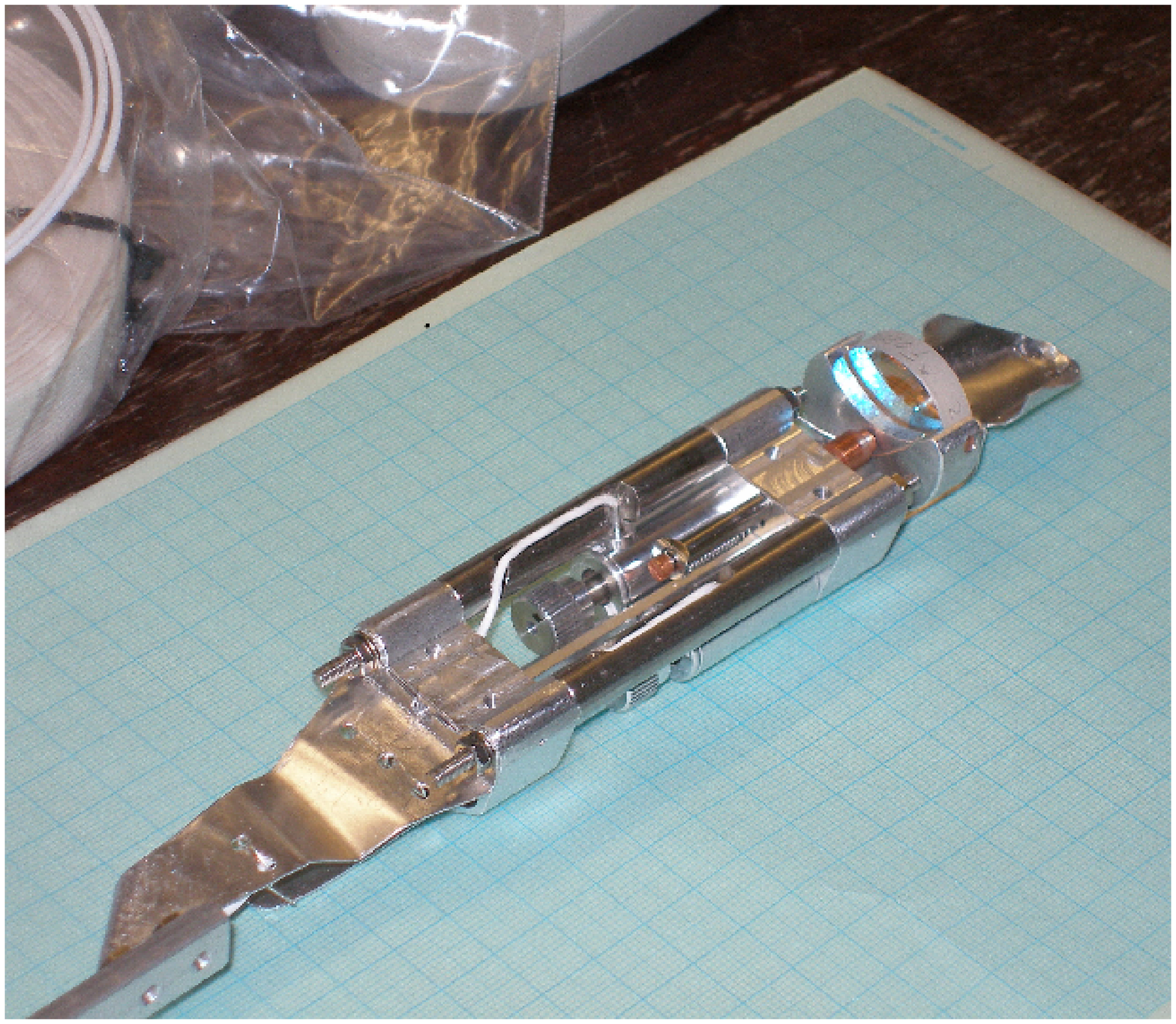}
\begin{picture}(10,0,1)
\put(8.5,4.3){\Large \bf (a)}
\put(6.5,4.6){\Large \bf (b)}
\put(1.4,3.6){\Large \bf (c)}
\end{picture}
\vspace*{-3mm}
\caption{Left: Photo of the ALPS laser bench with the setup for the resonant cavity. On the right one sees the LIGO-type infrared laser (a), right to middle the resonant second harmonic generation (b) and on the left the outer end of the resonant ALPS cavity with the vacuum vessel (c) containing the coupling mirror. Right: Photo of the ALPS end mirror setup with two Squiggle motors.  See the text for details.
\label{fig:ALPS_laser}}
\end{figure}

The most ambitious and crucial part of the ALPS experiment is the laser system and the resonant photon generation, 
cf.~Fig.~\ref{fig:ALPSsetup} and~Fig.~\ref{fig:ALPS_laser}. ALPS is the first experiment, which successfully exploit a large-scale optical resonator for WISP searches. The main parts and their basic functionality will be explained in the following, for more details refer to \cite{ALPSNIM} and reference therein. 
As light source a LIGO-type single frequency MOPA\footnote{Master-Oscillator Power Amplifier} laser system is used, producing 35~W of 1064~nm laser light. A piezo-electric transducer installed on the generic non-planar ring oscillator (NPRO) laser crystal allows for a frequency shift of $\pm 100$~MHz with a response bandwidth of 100~kHz.
In order to optimize the detection efficiency\footnote{An efficient detection of green light (532~nm) is much easier than for
infrared (1064~nm).} the frequency of the beam is doubled with a non-linear PPKTP\footnote{Periodically Poled KTiOPO$_4$} crystal. 
This beam is redirected into the production vacuum tube in which photon-WISP conversions could occur. 
An optical resonator inside this pipe is used to buildup the laser power, enhancing proportionally the WISP flux.
The difference between the laser frequency and the actual resonance frequency of the cavity is determined via 
a sideband modulation spectroscopy technique\footnote{Pound-Drever-Hall scheme}. This differential input  is used 
by an electronic feedback control loop, which adopts the frequency of the infrared MOPA laser in order
to lock the cavity. Variations of the resonator frequency are dominated 
by length fluctuations of the setup.

The gain of the resonator, given by the power build-up $PB$, which is the  ratio of the laser power inside the resonator
to the incident laser power, is limited by the internal losses of the cavity.
In the first setup of an optical resonator at ALPS \cite{ALPSNIM}, the mirrors of the optical cavity were placed outside the production side vacuum tube, so that the green laser beam had to traverse two glass windows two times in one resonator round trip. 
Absorption and scattering in these windows (although AR-coated) limited the achievable power build-up. 
By placing the mirrors inside the vacuum, the internal losses of the production resonator in the current setup were reduced by roughly an order of magnitude, boosting the power build-up by the same factor to $PB \approx 300$.
The green laser light is directed through the entrance window of the cylindrical vacuum vessel (cf.~Fig.~\ref{fig:ALPS_laser}) onto the coupling mirror (CM) 
of the cavity. 
Using two UHV Picomotors  the CM mount is adjustable from the outside. 
For EM a special mirror holder was designed which is non-magnetic and suitable for high vacuum.  
With two  Squiggle motors (cf.~Fig.~\ref{fig:ALPS_laser}) the mirror mount in the holder can be tilted remotely around two axes perpendicular to the beam and to each other. 
%

\subsubsection{Resonant Second Harmonic Generation}
The efficiency of a single pass red-green conversion is even under optimized conditions just around 2\% for a continous beam.
In order to increase the available 532~nm laser power, a folded ring shaped resonator was build around the nonlinear PPKTP crystal used for the second harmonic generation (SHG), cf.~Fig.~\ref{fig:ALPSsetup} and~Fig.~\ref{fig:ALPS_laser}. 
The length is constantly changed by an electronic feed-back loop in order to keep it resonant with the incident infrared laser light. This resonant SHG scheme pushes the conversion efficiency to 50\%. 
In the stable long-term operation the resonant SHG emits up to 5~W  of 532~nm laser light from an incident power of 10~W at 1064~nm. 
During the measurement period in the year 2009 the laser power feed into the cavity was kept below 5~W to minimize potential degradation of the cavity mirrors \cite{ALPSResult}, resulting in a continuously circulating power inside the ALPS production region of around 1.2~kW.

\subsection{Tuning of the Refractive Index}
ALPS also exploits successfully a new method to cover the gaps in the sensitivity for higher masses, where the ALP wave runs out of phase w.r.t.~the phase of the laser beam, cf.~Fig.~\ref{fig:ALPSprospects}.
Introducing Ar gas at a pressure of 0.18~mbar changes the photon momentum and tunes therefore the refraction index. In the ALPS
setup the $\gamma-$ALP relative phase velocity increases thereby to have an extra half oscillation length. Even if the sensitivity is lowered compared to vacuum conditions this helps to cover the high mass gaps, cf.~Fig.~\ref{fig:ALPSresultalps}.

\subsection{ALPS Result}
ALPS took around 50 data sets (1~h frames) under different experimental conditions: with magnet on or off,  laser polarization parallel or perpendicular to the magnetic field and different gas pressures. Details on the methodology and analysis are described in \cite{ALPSNIM, ALPSResult}. From the non observation of any WISP signal a 95~\% confidence level on the conversion probabilty was obtained, ranging between $ P_{\gamma \rightarrow \phi \rightarrow \gamma} = 1 ... 10 \times 10^{-25}$ for the different experimental setups.
\begin{figure}
\includegraphics[height=60mm]{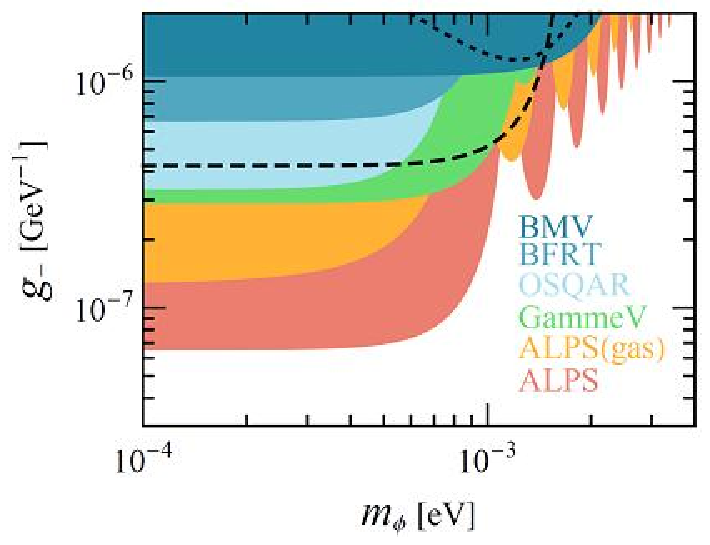}
\includegraphics[height=60mm]{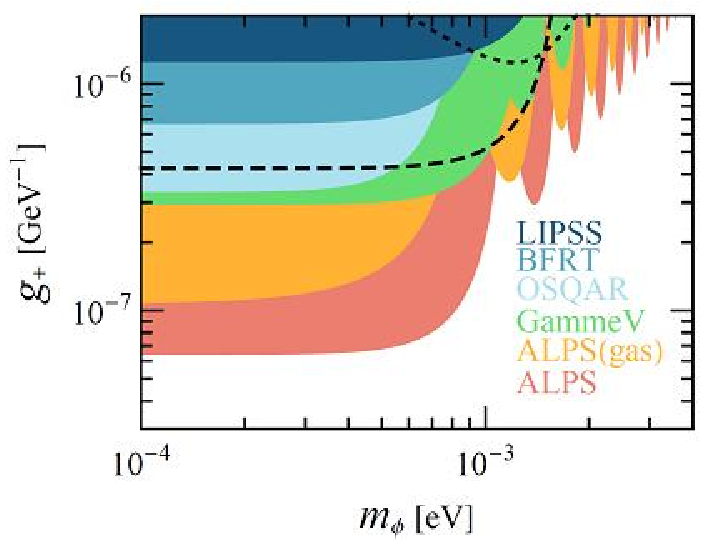}
\caption{Exclusion limit (95\% C.L.) for pseudoscalar (left)  and scalar (right) axion-like particles obtained by the ALPS experiment from vacumm and gas runs together with the results from various other LSW experiments \cite{ALPSResult}, 
see the text for details.
Dashed and dotted lines show the bounds derived form the PVLAS measurement on ALP induced dichroism and birefringence \cite{PVLAS}.
\label{fig:ALPSresultalps}}
\end{figure}
Fig.~\ref{fig:ALPSresultalps} shows the ALPS results for pseudoscalar  and scalar  axion-like particles
together 
with the results obtained from BMV \cite{BMV},  BFRT \cite{BFRT},  GammeV \cite{GammeV},  LIPSS \cite{LIPSS} and OSQAR \cite{OSQAR}. The gaps at higher masses are covered by the ALPS gas runs as described above. ALPS provide now the most stringent laboratory bounds on ALPs in the sub-eV mass range.

\begin{figure}
\includegraphics[height=60mm]{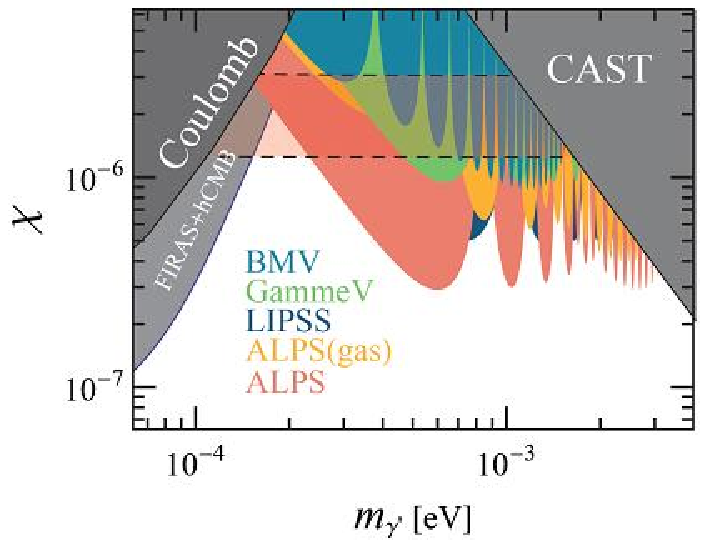}
\includegraphics[height=60mm]{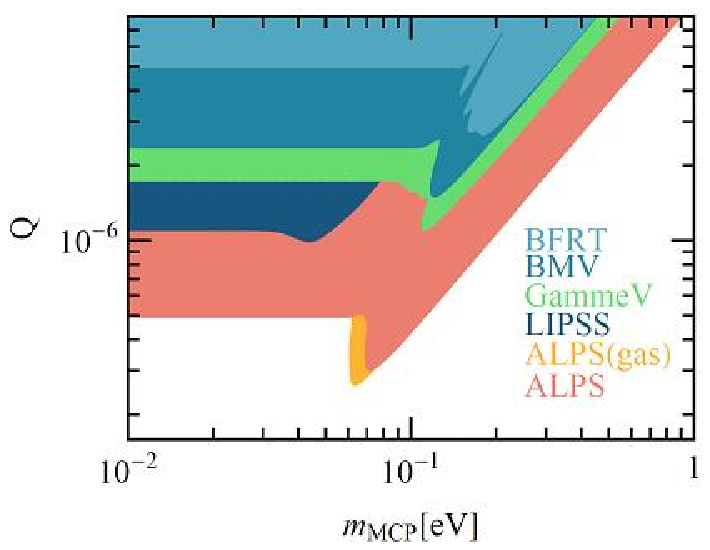}
\caption{ALPS exclusion limit (95\% C.L.) for hidden photons (left) and minicharged particles (right) together with the results  from various other experiments \cite{ALPSResult}.
\label{fig:ALPSresult_hs_mcp}}
\end{figure}

Also for hidden photon and minicharged particle search ALPS provides now the most stringent laboratory bounds on their existence, cf.~Fig.~\ref{fig:ALPSresult_hs_mcp}. The ALPS LSW results on hidden photon search fills the gap 
between lab searches for deviations from Coulomb's law and astrophysical bounds. Remarkable, with the achieved sensitivity ALPS almost completely rules out the hint of WMAP and large-scale-structure probes with non-standard radiation density contribution due to hidden photons, cf.~\cite{ALPSResult} and references therein.

\section{Prospects of Direct WISP Search Experiments}
Further upgrades and plans toward large scale LSW experiments are aiming to surpass present astrophysics limits on the coupling of ALPs to photon. This requires a sensitivity in the photon-ALP coupling of $g<10^{-10}\rm{GeV}^{-1}$, an improvement of 3 orders of magnitude with respect to the actual ALPS results. Table~\ref{tab:alp_coupling} summarizes the dependence of the sensitivity in $g$ on experimental parameters together with possible improvements. 

\begin{table}[t]
\renewcommand{\arraystretch}{1.4}  
\caption{Dependence of the photon-ALP coupling on experimental parameter together with the gain in sensitivity of future experiments with respect to the actual ALPS setup, see text for details.\label{tab:alp_coupling}}
\vspace{0.4cm}
\begin{center}
\begin{tabular}{|l|c|c|c|c|}
\hline
Parameter & g dependence &  ALPS &  future exp.& gain\\
\hline
Magnetic field & $ g \propto  BL^{-1}$ & $ BL = 23$~Tm & $BL = 300$~Tm & 13
\\ 
\hline
Laser power & $g \propto  P^{-\frac{1}{4}}$ & $ P=1$~kW & $ P=100$~kW & 3.2 \\ \hline

Detector sensitivity & $g \propto  \epsilon^{\frac{1}{4}} $ & $ \epsilon=2$~mHz & $ \epsilon=0.02$~mHz & 3.2 \\ \hline

Measurement time\footnotemark & $g \propto  t^{-\frac{1}{8}} $& $ t=10$~h & $ t=1000$~h & 1.8 \\ \hline

Resonant regeneration & $g \propto  PB^{-\frac{1}{4}} $& $ PB=1$ & $ PB=10000$~h & 10 \\ \hline

\end{tabular}
\end{center}
\end{table}

\subsubsection{Magnet}
The sensitivity in $g$ improves linearly with the magnetic field strength and length.
Instead of half an HERA dipole magnet with $BL \approx 23$~Tm as used within the actual ALPS setup for the WISP generation 
and for the reconversion to photons one may use e.g.~up to six HERA dipoles on each side providing about 280~Tm.
This would improve the sensitivity by more than one order of magnitude. 
Alternatively two plus two LHC magnets, which are the most powerful existing magnets for this purpose, could be used, providing
$BL \approx 300$~Tm. For ultimate experiments one may even think of 4 or 6 LHC magnets on each side or at some point even more powerful magnets may become available. 
\footnotetext{For detectors limited in their sensitivity by background counting rates.}

\subsubsection{Laser Power and Detector}
It looks feasible to increase in further experiments the incident laser power to the cavity by a factor of 10 
and to improve the power build-up in the resonant cavity by an additional factor of 10.
Therefore a laser power of around 100~kW in the generation part seems to be realistic. 

The detector sensitivity for the actual ALSP setup is limited by dark current and read-out noise. 
Single photon counting techniques, e.g.~with cryogenic transition edge sensors \cite{TES} 
may provide a factor up to 100 improvement in the sensitivity. However, this is to be worked out in a dedicated R\&D programme.

This results in an additional order of magnitude improvement in the sensitivity in $g$, but still not enough to surpass astrophysical limits. More statistics, i.e.~longer measurement time,
will not really help, even a 100 times longer data taking period provides less than a factor of two improvement.

\subsubsection{New Idea - Resonant Regeneration}
An old idea from the 1990's was recently rediscovered, to set up similar to the generation part
an additional optical cavity for resonant axion photon regeneration, 
which enhances the small electromagnetic photon component of a 
potential WISP wave behind the wall \cite{RR_Hoogeveen, resALPreg}. The technical details are rather challenging, 
e.g.~one can obviously not use laser light of the same wavelength for locking and for the WISP production.
There are different proposals under discussions \cite{resALPreg, ALPSresreg}. The ALPS experiment intend to use 1064~nm laser light for the WISP production and frequency doubled laser light with 532~nm for locking of the regeneration cavity.
A power build of $PB \approx 10000$ seems to be possible, which would increase the sensitivity to $g$ by another order of magnitude.

\begin{figure}
\centerline{\includegraphics[height=60mm]{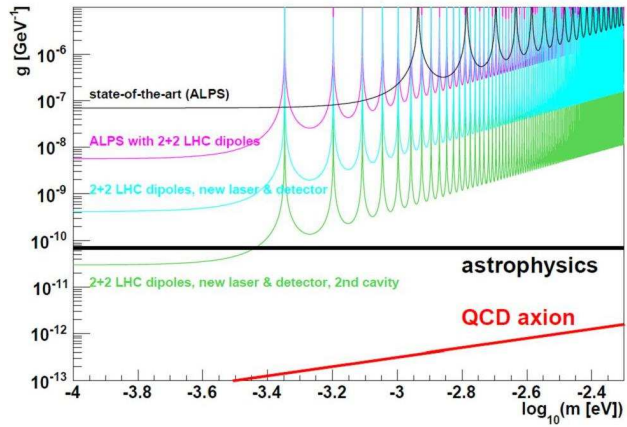}}
\caption{Prospects of ALPS II: expected gain in sensitivity to $g$, cf.~text. 
\label{fig:ALPSprospects}}
\end{figure}

Fig.~\ref{fig:ALPSprospects} summarizes the  gain in sensitivity to the photon-ALP coupling for the various improvements described above.


\section{Summary and Outlook}
The low energy frontier is a promising complement in particle physics to the high-energy frontier.
Worldwide interest and activities in laboratory  experiments for WISP search grew up within the last years and   
complement astrophysics searches.
``Light shining through a wall'' experiments are an intriguing  simple idea for direct WISP search. 
They demonstrated successfully their capability and open a new window to explore hidden worlds. 
The ALPS experiment at DESY provides now the most stringent laboratory constraints on the existence of WISPs.
This success is based on close collaboration between particle physicist and laser physicists from the gravitational wave detector community and the infrastructure and support of a high-energy physics laboratory. 
Based on this experience a detailed planing of future large scale LSW experiments which improves the sensitivity by orders of magnitudes has started. It looks very promising to surpass present day limits from astrophysics, but finding the QCD axion remains a very challenging task.

\end{document}